\documentclass[prd
,preprint
,superscriptaddress,showpacs,nofootinbib,%
tightenlines]{revtex4}
\usepackage{epsfig}
\usepackage{amssymb,amsmath}
\usepackage{ulem}
\newcommand{\beq}{\begin{equation}}
\newcommand{\eeq}{\end{equation}}
\newcommand{\beqa}{\begin{eqnarray}}
\newcommand{\eeqa}{\end{eqnarray}}

\newcommand{\ben}{\begin{displaymath}}
\newcommand{\een}{\end{displaymath}}
\newcommand{\be}{\begin{equation}}
\newcommand{\ee}{\end{equation}}
\newcommand{\bea}{\begin{eqnarray}}
\newcommand{\eea}{\end{eqnarray}}
\begin{document}
\title{
Deuteron electromagnetic form factors in a renormalizable formulation of chiral effective field theory}
\author{E.~Epelbaum}
\affiliation
{Institut f\" ur Theoretische Physik II, Fakult\" at f\" ur Physik und Astronomie, \\ Ruhr-Universit\" at Bochum 44780 Bochum, Germany}
\author{A.~M.~Gasparyan}
\affiliation
{Institut f\" ur Theoretische Physik II, Fakult\" at f\" ur Physik und Astronomie, \\ Ruhr-Universit\" at Bochum 44780 Bochum, Germany}
\affiliation
{SSC RF ITEP, Bolshaya Cheremushkinskaya 25, 117218 Moscow, Russia}
\author{J.~Gegelia}
\affiliation
{Institut f\" ur Theoretische Physik II, Fakult\" at f\" ur Physik und Astronomie, \\ Ruhr-Universit\" at Bochum 44780 Bochum, Germany}
\affiliation
{Tbilisi State University, 0186 Tbilisi, Georgia}
\author{M.~R.~Schindler}
\affiliation
{Department of Physics and Astronomy\\
University of South Carolina,
Columbia, SC 29208
USA}
 \date{28 November, 2013}
\begin{abstract}

We calculate the deuteron electromagnetic form factors in a
modified version of Weinberg's chiral effective field theory
approach to the two-nucleon system. We derive renormalizable
integral equations for the deuteron without partial wave
decomposition. Deuteron form factors are extracted by applying
the Lehmann-Symanzik-Zimmermann reduction formalism to the three-point correlation function of
deuteron interpolating fields and the electromagnetic current operator. Numerical results of
a leading-order calculation with removed cutoff regularization agree well with experimental data.

\end{abstract}

\maketitle

\section{\label{introduction}Introduction}

The seminal papers by Weinberg on chiral effective field theory (EFT)
of nuclear forces \cite{Weinberg:rz,Weinberg:um} have triggered an
intense activity starting with Ref.~\cite{Ordonez:1992xp}. For recent
reviews see e.g. Refs.~\cite{Epelbaum:2008ga,Machleidt:2011zz}.
One of the most discussed aspects of the application of chiral
effective field theory to two- and few-body problems is related to the
question of how to properly renormalize the resulting integral equations.
A new framework to solve this problem was proposed in
Ref.~\cite{Epelbaum:2012ua}, which is based on the manifestly
Lorentz-invariant effective Lagrangian and time-ordered perturbation
theory.
Within this scheme the leading-order (LO) nucleon-nucleon scattering amplitude is obtained by solving an integral
equation (known as the Kadyshevsky equation \cite{kadyshevsky}), and corrections are calculated perturbatively.
The LO equation is perturbatively renormalizable due to the milder
ultraviolet behavior of the two-nucleon propagator compared to the
standard heavy-baryon formalism. Partial wave projected equations have unique
solutions except for the $^3P_0$ wave which requires a special
treatment. In the present study we calculate the electromagnetic form
factors of the deuteron, and therefore the issue of the $^3P_0$ wave is not
relevant  here.

The electromagnetic structure of the deuteron has been extensively
analyzed in the EFT framework using various approaches. In particular, Kaplan, Savage and
Wise \cite{Kaplan:1998sz}  calculated the electromagnetic form factors of the deuteron
up to next-to-leading order (NLO) in a framework based on a
perturbative treatment of potential pions and found good agreement
with data up to momentum transfers of the order of $q\sim
400\,\text{MeV}$. Shortly thereafter a number of calculations based on
Weinberg's approach (or variations thereof) with nonperturbative pions
have been performed at various orders in the chiral expansion
\cite{Phillips:1999am,Walzl:2001vb,Phillips:2003jz,Phillips:2006im,Kolling:2012cs},
see also Refs.~\cite{Ordonez:1995rz,Park:1997kp} for pioneering
quantitative studies of nucleon-nucleon scattering in this framework
and Refs.~\cite{Epelbaum:2008ga,Machleidt:2011zz} for recent review articles.
Generally, after employing factorization in order to account for
single-nucleon electromagnetic structure, a good description of the
deuteron form factors up to rather high values of the momentum
transfer was reported in all these calculations provided
the isoscalar single-nucleon form factors are accurately described.\footnote{Note that the
factorization amounts to taking into account higher-order
  terms in the chiral expansion of the single-nucleon current
  operator.} Two-body currents, worked out at
leading loop order in the heavy-baryon formulation of chiral EFT by
Park et al.~\cite{Park:1995pn} and re-derived recently by the JLab-Pisa \cite{Pastore:2008ui,Pastore:2009is,Pastore:2011ip}  and
Bochum-Bonn groups \cite{Kolling:2009iq,Kolling:2011mt}, are mainly of
isovector type and thus play only a minor role for the
deuteron. Further applications of the exchange currents to the
electromagnetic structure and reactions in two- and three-nucleon
systems are reported in Refs.~\cite{Rozpedzik:2011cx,Piarulli:2012bn}.
With the exception of Ref.~\cite{Kaplan:1998sz}, which makes use of dimensional
regularization, all these calculations employ a finite
ultraviolet cutoff $\Lambda$ chosen to be smaller or of the order of
the rho-meson mass. Much larger cutoff values in the range of $\Lambda
\lesssim 4\,\text{GeV}$ are considered in
Ref.~\cite{Valderrama:2007ja} together with the leading- and
next-to-next-to-leading order chiral wave functions.

In the present work, we extend our recently suggested renormalizable formulation of
nuclear chiral EFT with non-perturbative pions \cite{Epelbaum:2012ua} to calculate the
electromagnetic form factors of the deuteron at lowest
order.  Similarly to Ref.~\cite{Fachruddin:2001sb}, we derive
a system of integral equations for the deuteron without making use of partial wave decomposition.
The crucial new feature of our framework is its explicit
renormalizability in spite of the non-perturbative treatment of the
one-pion exchange (OPE) potential. This allows us to safely  take
the cutoff parameter to infinity after subtracting the ultraviolet
divergences. Our paper is organized as follows. In section \ref{sec2}
we briefly outline a general  formalism to calculate the form factors of the
deuteron in quantum field theory. The integral equations for the
deuteron interpolating field interacting with a pair of nucleons are
worked out in section \ref{sec3}. Finally, a discussion and summary of
the obtained results are given in section \ref{sec4}.

\section{The deuteron form factors}
\label{sec2}

We use the conventions and notations of Ref.~\cite{Kaplan:1998sz}.
The deuteron is characterized by its momentum $P^\mu$ and polarization
$\epsilon^\mu$.
The polarization vectors can be expanded in terms of basis vectors
$\epsilon^\mu_i$ ($i=1,2,3$), which satisfy the conditions
\begin{equation}
P_\mu{\bf\epsilon}^\mu_i=0, \ \
\epsilon^*_{i\mu}\epsilon^\mu_j=-\delta_{ij}, \ \
\sum_{i=1}^3\epsilon^{*\mu}_{i}\epsilon^\nu_i= \frac{P^\mu P^\nu}{M_d^2}-g^{\mu\nu},
\label{polvector}
\end{equation}
where $M_d=2\,m-B$ is the deuteron mass, $B$ its binding
energy, and $m$ the nucleon mass. We choose these polarization vectors so that in the rest frame
of the deuteron $\epsilon^\mu_i=\delta^\mu_i$ and denote deuteron
states with $|{\bf P},i\rangle$ ($\equiv|{\bf
  P},\epsilon^\mu_i\rangle$).  These states satisfy the normalization condition
\begin{equation}
\langle {\bf P}\,',j|{\bf P},i\rangle=\frac{P^0}{M_d}\,(2\,\pi)^3\,\delta^3({\bf P}-{\bf P}\,')\delta_{ij}.
\label{normalization}
\end{equation}

The matrix element of the electromagnetic current operator to leading
order in a non-relativistic expansion can be parameterized as
\begin{eqnarray}
\langle {\bf P}\,',j|J^0_{em}|{\bf P},i\rangle &=& e \left[
  F_C(q^2)\,\delta_{ij} +\frac{1}{2 M_d^2}\,F_Q(q^2)\left({\bf
      q}_i{\bf q}_j-\frac{1}{3}\,q ^2 \delta_{ij}\right)\right]\,,\nonumber\\
\langle {\bf P}\,',j|J^k_{em}|{\bf P},i\rangle &=& \frac{e}{2 M_d}
\Biggl[ F_C(q^2)\,\delta_{ij} \left({\bf P}+{\bf P}\,'\right)^k
+F_M(q^2)\left( \delta_j^k {\bf q}_i-\delta_i^k {\bf
    q}_j\right)\nonumber\\
&+& \frac{1}{2 M_d^2}\,F_Q(q^2)\left({\bf q}_i{\bf
    q}_j-\frac{1}{3}\,q^2 \delta_{ij}\right)\left({\bf P}+{\bf
    P}\,'\right)^k\Biggr],
\label{DFF}
\end{eqnarray}
where ${\bf q}={\bf P}\,'-{\bf P}$ is the transferred momentum and
$q=|{\bf q}|$. The form factors are normalized as follows (see
e.g. Ref.~\cite{Zuilhof:1980ae}),
\begin{equation}
F_C(0)=1, \ \ \frac{e}{2 M_d}\,F_M(0)=\mu_M, \ \ \frac{1}{M_d^2}\,F_Q(0)=\mu_Q,
\label{ffnormalization}
\end{equation}
with $\mu_M=0.857 4 (e/(2 m))$  being the deuteron magnetic moment \cite{Mohr:2012tt}
and $\mu_Q=0.2859\ {\rm fm}^2$ its quadrupole moment \cite{Bishop:1979zz,Ericson:1982ei}.
It is also common to parameterize the matrix elements of the current
operators in terms of the three form factors $G_C$, $G_M$, and $G_Q$, where
\be
G_C(q^2)=F_C(q^2),\quad G_M(q^2)=F_M(q^2),\quad G_Q(q^2)=\frac{1}{M_d^2}F_Q(q^2).
\ee

\medskip

We follow Ref.~\cite{Kaplan:1998sz} and define the deuteron interpolating field as
\begin{equation}
{\cal D}_i\equiv N^T \mathcal{P}_i N=\sum_{\alpha,\beta,a,b=1}^2 N_{\alpha,a} \mathcal{P}_{i,a,b}^{\alpha\beta} N_{\beta,b}
, \ \ \ \mathcal{P}_i\equiv \frac{1}{\sqrt{8}}\,{\sigma_2\sigma_i \tau_2},
\label{interfield}
\end{equation}
where $\alpha, \,\beta$ and $a, \,b$ are spin and isospin indices, respectively.
This choice is made for convenience, and observables do not depend on the particular form of the interpolating field.
The full propagator $G_{\cal D}$ is given by the time-ordered product of two interpolating fields,
\begin{equation}
G_{\cal D}(P) \delta_{ij}=\int d^4 x e^{-i P x} \langle 0|T\left[{\cal D}_j^\dagger(x) {\cal D}_i(0)\right]|0\rangle = \delta_{ij}\,\frac{i\,{\cal Z}(P^2)}{P^2-M_d^2+i\,\epsilon}.
\label{propagatorD}
\end{equation}

The electromagnetic form factors of the deuteron are related to the three-point function of the electromagnetic current operator $J_{em}^\mu$ and two interpolating fields,
\begin{equation}
G_{ij}^\mu(P,P\,')=\int d^4 x d^4 y\, e^{-i P y}  e^{i P\,' x}\langle 0|T\left[{\cal D}_j^\dagger(x) J_{em}^\mu(0)\,{\cal D}_i(y)\right]|0\rangle ,
\label{VertexD}
\end{equation}
through the Lehmann-Symanzik-Zimmermann (LSZ) reduction formula \cite{Kaplan:1998sz},
\begin{eqnarray}
\langle {\bf p\,'},j|J_{em}^\mu|{\bf p},i\rangle &=& Z \left[ G^{-1}(P)G^{-1}(P\,') G_{ij}^\mu(P,P\,')\right]_{P^2,P\,'^2\to M_D^2}\nonumber\\
&=&-\frac{1}{Z}\,\left[\left(P^2-M_d^2\right)\left(P\,'^2-M_d^2\right) G_{ij}^\mu(P,P\,')\right]_{P^2,P\,'^2\to M_d^2},
\label{LSZ}
\end{eqnarray}
where $Z={\cal Z}(M_d^2)/(2 M_d)$ is the residue of the propagator. In
other words, the form factors can be extracted from the residue of the
double pole of the vertex function $G_{ij}^\mu(P,P\,')$.

\begin{figure}
\epsfig{file=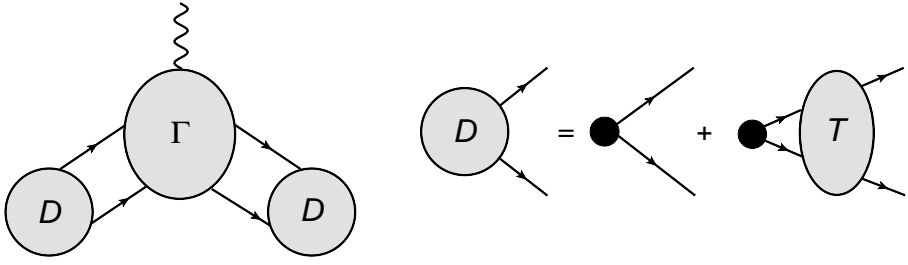, width=12cm}
\caption[]{\label{vertex} Vertex function of the electromagnetic current and two interpolating fields of the deuteron. The circle with $\Gamma$ stands for the two-nucleon irreducible part of the vertex function, $D$ is the amplitude of the deuteron interpolating field interacting with a pair of nucleon fields. 
The black circles represent the interaction vertices of the interpolating field with a pair of nucleons, solid lines represent the nucleons and the waved line corresponds to the electromagnetic current.}
\end{figure}

The vertex function can be represented diagrammatically as shown in
Fig.~\ref{vertex}. It consists of the two-nucleon irreducible part
$\Gamma$ connected to two nucleon-nucleon scattering amplitudes
contracted to vertices corresponding to interpolating operators,
denoted by $D$. For our LO calculations,  we take the irreducible term in the form of the one-nucleon current,
\begin{equation}
\Gamma^\mu(p',p) = \frac{F_1(q^2)}{2\,m}\,\left(p+p'\right)^\mu +i\,\frac{F_1(q^2)+F_2(q^2)}{2\,m}\ \delta^\mu_i \epsilon^{ijk} q^j \sigma^k,
\label{GammaTree}
\end{equation}
where $F_1$ and $F_2$ are the electromagnetic form factors of the nucleon, $q_\mu=p^\prime_\mu-p_\mu$, with $p_\mu$ and $p^\prime_\mu$ the four momenta of incoming and outgoing nucleons, respectively.

\section{The deuteron equation}
\label{sec3}

In the modified EFT approach of Ref.~\cite{Epelbaum:2012ua} the LO NN scattering amplitude is obtained by solving
the integral equation
\begin{equation}
{T \left(
{\bf p\,'},{\bf p}\right)}{=}{ V \left(
{\bf p\,'},{\bf p}\right) - m^2 \int \frac{d^3 {\bf k}}{(2\,\pi)^3}} {\frac{V \left(
{\bf p\,'},{\bf k}\right) \, T \left(
{\bf k},{\bf p}\right)}{\omega_k^2\, \left(
E-2 \,\omega_k+i\,\epsilon\right)},}
\label{kad}
\end{equation}
where $E=2 \sqrt{{\bf p}^2+m^2}$   denotes the energy of two incoming nucleons in the
center of mass frame and $\omega_k=\sqrt{{\bf k}^2+m^2}$.
The LO NN potential can be taken in the usual form
\begin{equation}
V_0\left(
{\bf p\,'},{\bf p}\right) = C_S+C_T\, {\bf\sigma_1}\cdot{\bf\sigma_2}
-\frac{g_A^2}{4\,F^2}\ {\bf \tau_1}\cdot{\bf\tau_2} \
\frac{{\bf\sigma_1}\cdot ({\bf p}\,'-{\bf p})\,{\bf\sigma_2}\cdot ({\bf p}\,'-{\bf p})}{({\bf p}\,'-{\bf p})^2+M_\pi^2}.
\label{LOV}
\end{equation}
By parameterizing the potential and
the scattering amplitude as (here we indicate explicitly the spin
indices  omitted in Eq.~(\ref{kad}))
\begin{equation}
\begin{split}
V_{\alpha\beta,\, \gamma\delta}\left( {\bf p'} ,{\bf p}\right) & =  v^0
\left( {\bf p'} ,{\bf p} \right)
\delta_{\alpha\gamma}\delta_{\beta\delta}+v^1_a \left( {\bf p'} ,{\bf p}
\right) \,\left(\sigma^a_{\alpha\gamma}\delta_{\beta\delta} +
\delta_{\alpha\gamma}\sigma^a_{\beta\delta}\right) +v^2_{ab} \left( {\bf
p'} ,{\bf p}\right)
\,\sigma^a_{\alpha\gamma}\sigma^b_{\beta\delta}\,,\\
T_{\alpha\beta,\, \gamma\delta}\left( {\bf p'} ,{\bf p}\right) & =  t^0
\left( {\bf p'} ,{\bf p} \right)
\delta_{\alpha\gamma}\delta_{\beta\delta}+t^1_a \left( {\bf p'} ,{\bf p}
\right) \,\left(\sigma^a_{\alpha\gamma}\delta_{\beta\delta} +
\delta_{\alpha\gamma}\sigma^a_{\beta\delta}\right) +t^2_{ab} \left( {\bf
p'} ,{\bf p}\right) \,\sigma^a_{\alpha\gamma}\sigma^b_{\beta\delta}\,,
\end{split}
\label{tpar}
\end{equation}
substituting in Eq.~(\ref{kad}),
simplifying the Pauli matrices, and equating the coefficients of
equal structures we arrive at the following system of equations:
\begin{equation}
t_{ab}({\bf p'},{\bf p})= v_{ab}({\bf p'},{\bf p}) - m^2 \int \frac{d^3{\bf k}}{(2\,\pi)^3} \,
W_{ab,xy}({\bf p'},{\bf k})\,G({\bf k})\,  t_{xy}({\bf k},{\bf p})\,,
\label{systemofequations}
\end{equation}
where
\begin{eqnarray}
t_{ab} & = & \left(  t^0, \;   t^1_a , \;
    t^2_{ab} \right)^{\rm T} ,
\nonumber\\[5pt]
v_{ab} & = & \left(  v^0, \;   v^1_a , \;
    v^2_{ab} \right)^{\rm T} ,
\nonumber\\[5pt]
W_{ab,xy} & = & \left(
           \begin{array}{ccc}
             v^0, & 2\,v^1_x, & v^2_{xy} \\
             v^1_a, & v^0 \delta_{a x}+i\,\epsilon^{amx}v^1_m + v^2_{ax}, & \delta_{ax} v^1_y +i\,\epsilon^{amx} v^2_{my} \\
             v^2_{ab}, & W_{32}, &  W_{33}\\
           \end{array}
         \right)\,,\nonumber\\[5pt]
W_{32} & = &
v^1_a\delta_{bx}+v^1_b\delta_{ax}+i\,\epsilon^{mxa}v^2_{mb}+
i\,\epsilon^{mxb}v^2_{ma},\nonumber\\[5pt]
W_{33} & = & v^0\delta_{ax}\delta_{by}
             -i\,\epsilon^{axd}\delta_{by}v^1_d-i\,\epsilon^{byd}\delta_{ax}v^1_d-\epsilon^{mxa}\epsilon^{nyb}v^2_{mn}\,,
\nonumber\\[5pt]
G({\bf k}) & = & {\frac{1}{\omega_k^2\, \left(
E-2 \,\omega_k +i\,\epsilon\right)}}.
\end{eqnarray}

The amplitude of the deuteron interpolating field interacting with a pair of nucleon fields in the rest frame of the deuteron is given by
\begin{equation}
D_j({\bf p\,'}) = \mathcal{P}_j + m^2\int \frac{d^3{\bf k}}{(2\,\pi)^3} \,
\frac{T({\bf p'},{\bf k})\,\mathcal{P}_j}{ \omega_k^2\, \left(
E-2 \,\omega_k+i\,\epsilon\right)}\,,
\label{TD}
\end{equation}
where the LO NN scattering amplitude is obtained by solving Eq.~(\ref{kad}).
It is convenient to parameterize the amplitude $D$ in terms of two
structure functions $\Delta_1$ and $\Delta_2$ as
\begin{equation}
D_{j}\left( {\bf p'}\right) = \Delta_1({\bf p'}^2) \mathcal{P}_j + {\bf p'}_a {\bf p'}_b\, \Delta_2({\bf p'}^2)\,\sigma^a \mathcal{P}_j\left(\sigma^b\right)^T,
\label{Dpar}
\end{equation}
where isospin indices and terms resulting from anti-symmetrization are
not shown explicitly. Notice that the structure functions $\Delta_1$ and $\Delta_2$
can be easily related to the $S$- and $D$-state components of the deuteron wave function, see also Ref.~\cite{Fachruddin:2001sb}.
To derive equations for the structure functions $\Delta_i$ of the
projected amplitude we parameterize the NN potential as
\begin{eqnarray}
v^0({\bf p'},{\bf p}) & = & \nu_1({\bf p'},{\bf p})\,,\nonumber\\
v^1_a({\bf p'},{\bf p}) & = & i\,\epsilon^{abc} {\bf p}^b {\bf p'}^c
\nu_3({\bf p'},{\bf p})\,, \\
v^2_{ab}({\bf p'},{\bf p}) & = & \delta_{ab}\, \nu_2({\bf p'},{\bf p}) +{\bf p'}^a {\bf p'}^b\,
\nu_5({\bf p'},{\bf p})+ {\bf p}^a
{\bf p}^b\, \nu_6({\bf p'},{\bf p}) +  ({\bf p}^a
{\bf p'}^b+ {\bf p'}^a {\bf p}^b)\,\nu_4({\bf p'},{\bf p}),  \nonumber
\label{fversusampl}
\end{eqnarray}
where the $\nu_i({\bf p'},{\bf p})$ are scalar functions of ${\bf
p'}^2$, ${\bf p}^2$ and ${\bf p'}\cdot{\bf p}$. We then obtain the following system of integral equations:
\begin{equation}
\begin{split}
\Delta_1({\bf p}^2) & =  1+ m^2\int\frac{d^3{\bf k}}{(2\,\pi)^3}\, G({\bf k})
\biggl\{  \Delta_1({\bf k}^2) \left[\nu_1({\bf p},{\bf k})+\nu_2({\bf p},{\bf k})+C_1 \nu_6({\bf p},{\bf k})\right]   \\
&+ \Delta_2({\bf k}^2) \biggl[C_1\left(\nu_1({\bf p},{\bf k})+\nu_2({\bf p},{\bf k})\right)+2 ({\bf p}\cdot {\bf k}) \nu_3({\bf p},{\bf k})+2\, {\bf k}^2 ({\bf p}\cdot {\bf k})\, \nu_4({\bf p},{\bf k})          \\
&+  \left[({\bf p}\cdot {\bf k})^2-C_1 {\bf p}^2\right]\nu_5({\bf p},{\bf k})+({\bf k}^2)^2 \nu_6({\bf p},{\bf k})\biggr]\biggr\},    \\
\Delta_2({\bf p}^2) & =   m^2 \int \frac{d^3{\bf k}}{(2\,\pi)^3}\, G({\bf k})\,
\biggl\{\Delta_1({\bf k}^2) [2 B\, \nu_4({\bf p},{\bf k})+C_2 \nu_6({\bf p},{\bf k})+\nu_5({\bf p},{\bf k})]         \\
&+ \Delta_2({\bf k}^2)
\biggl[C_1 \nu_5({\bf p},{\bf k})
-2 B {\bf k}^2 \nu_3({\bf p},{\bf k})            \\
&+ C_2 \left(\nu_1({\bf p},{\bf k})
+ \nu_2({\bf p},{\bf k})+2\, ({\bf p}\cdot {\bf k})\, \nu_3({\bf p},{\bf k})\right)
\biggr]\biggr\},
\end{split}
\label{dsystem2}
\end{equation}
where  we have defined
\begin{equation}
B  \equiv  \frac{({\bf p}\cdot {\bf k})}{{\bf p}^2},\quad
C_1 \equiv  \frac{1}{2}\,\left[{\bf k}^2- \frac{({\bf p}\cdot {\bf k})^2}{{\bf p}^2}\right],\quad
C_2  \equiv  \frac{3\,({\bf p}\cdot {\bf k})^2-{\bf k}^2 {\bf p}^2}{2\,({\bf p}^2)^2}.
\label{BCDef}
\end{equation}
As the $\Delta_i$-functions depend only on ${\bf k}^2$, the
integration over angles can be carried out explicitly in
Eqs.~(\ref{dsystem2}) so that one is finally left with a system of two
one-dimensional integral equations which can be solved
numerically. The deuteron manifests itself as a pole at $P^2=E^2-{\bf
  0}^2=M_d^2$. 
Equations (\ref{dsystem2}) are divergent and require
regularization. Here, we employ cutoff regularization.
However, since all ultraviolet divergences can be absorbed into a redefinition of
the low-energy constant $C_{^3S_1}=C_S+C_T$,  we take the cutoff parameter $\Lambda$ to
infinity after renormalization.

The three-point function is given by
\begin{equation}
G^\mu_{ij}(P,P')= m^3 \int \frac{d^3 {\bf k}}{(2 \pi)^3} \frac{
  D^\dagger_{j,\alpha\beta}(P',{\bf
    k})\Gamma^\mu_{\alpha\beta,\alpha_1\beta_1}(P,P',{\bf k})
  D_{i,\alpha_1\beta_1}(P, {\bf k})}{\omega_{k}\,
  \omega_{k-\frac{q}{2}}\,\omega_{k+\frac{q}{2}}\left(E-\omega_{k}-
    \omega_{k-\frac{q}{2}}\right)\left(E-\omega_{k}-
    \omega_{k+\frac{q}{2}}\right)},
\label{TGT}
\end{equation}
where $D^\dagger_{j,\alpha\beta}(P',{\bf k}) $ and  $
D_{i,\alpha_1\beta_1}(P, {\bf k'})$ denote the amplitudes of the
deuteron interpolating field interacting with a pair of nucleons in a general frame. As
appropriate at LO, these quantities can be obtained from the
amplitudes calculated in the rest frame of the deuteron, see Eq.~(\ref{TD}), by means of a Galilean transformation.
We choose to work in the Breit frame, so that
$D^\dagger_j (P',{\bf k}) = D^\dagger_j ({\bf k} + {\bf q}/4)$ and
$D_i (P,{\bf k}) = D_i ({\bf k} - {\bf q}/4)$.

Using Eq.~(\ref{Dpar}), the deuteron full propagator $G_{\cal D}$ can be written as
\begin{equation}
G_{\cal D}\left(E,{\bf 0}\right)= m^2 \int \frac{d^3 {\bf
    k}}{(2\,\pi)^3} \frac{\Delta_1 \left({\bf k}^2\right)+\frac{{\bf
      k}^2}{3}\, \Delta_3 \left({\bf k}^2\right)}{\omega_k^2\, \left(
E-2\,\omega_k +i\,\epsilon\right)}+{\rm N.\, P.},
\label{2pointf}
\end{equation}
where ''N. P.'' stands for contributions which do not contain the
deuteron pole. We use  Eq.~(\ref{2pointf}) to calculate the residue
$Z$.

\section{Discussion and Summary}
\label{sec4}

\begin{figure*}
\centering
  \includegraphics[width=1.0\textwidth]{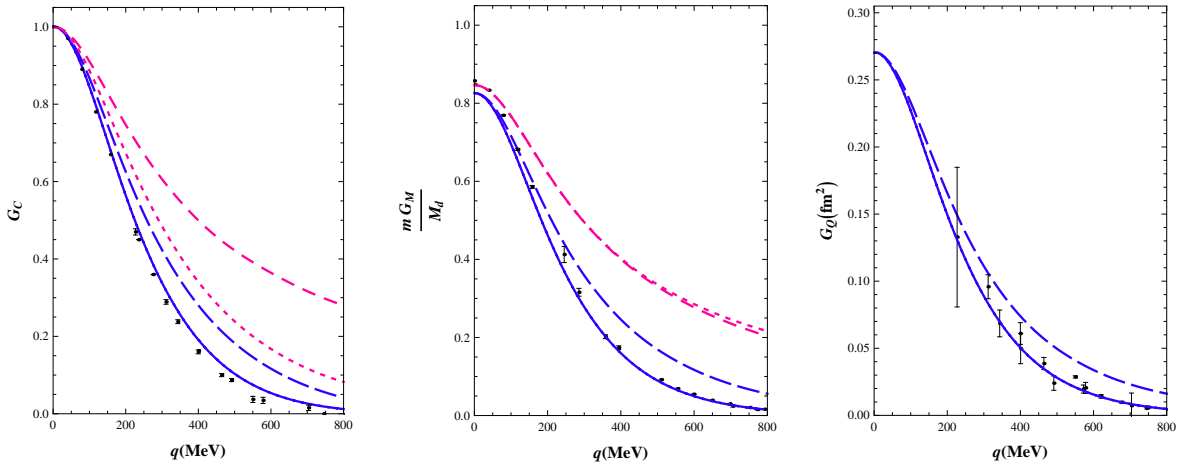}
\caption{LO EFT predictions for the Coulomb (left panel),  magnetic (middle
  panel) and quadrupole (right panel) form factors of the deuteron
  as a function of the momentum transfer $q$ in comparison with experimental
  data from Refs.~\cite{Abbott:2000ak,Nikolenko:2003zq}. Solid and long-dashed
  lines (short- and medium-dashed lines) show the predictions in the
  chiral (pionless) EFT with and without using phenomenological form factors of the
  nucleon, respectively.
}
\label{fig:2}
\end{figure*}

Using the formalism outlined above, we calculate the electromagnetic
form factors of the deuteron at LO by solving the integral
equations numerically.
We employ exact isospin symmetry as appropriate at LO
and use the following values for the low-energy constants entering the OPE potential:
\begin{equation}
M_\pi = 138\mbox{ MeV}, \quad 	F_\pi = 92.4\mbox{ MeV},  \quad g_A = 1.267\,.
\end{equation}
The low-energy constant  $C_{^3S_1}$ is fixed to reproduce the
experimental value of the deuteron binding energy of $B=2.22$ MeV.
The resulting description of  neutron-proton phase shifts and the quark mass
dependence of the S-wave scattering lengths and the deuteron binding
energy can be found in Refs.~\cite{Epelbaum:2012ua} and \cite{Epelbaum:2013ij},
respectively.\footnote{Note that in Ref.~\cite{Epelbaum:2012ua}, the low energy constant was
adjusted to the empirical phase shifts rather then to the deuteron binding energy. This leads, however,
only to a minor difference in the produced nucleon-nucleon scattering amplitude.}  In particular, the $^3S_1$
and $^3D_1$ phase shifts and the mixing angle $\epsilon_1$ turn out to
be reasonably well described (at least) up to laboratory energies of the order
of $E_{\rm lab} \sim 250$ MeV.  The corresponding parameter-free and
cutoff-independent  predictions for the electromagnetic form
factors of the deuteron are plotted in Fig.~\ref{fig:2}.
In all cases the obtained results agree reasonably well with
experimental data in the whole plotted range of the momentum transfer.
For low values of $q$ all three form factors are
accurately predicted at LO with the deviations increasing at high
momentum transfers and
reaching about $\sim 20 \%$  at $q = 200$ MeV.
Similarly to observations made in earlier studies based on
the non-relativistic approach
\cite{Phillips:2003jz,Phillips:2006im,Valderrama:2007ja}, the
deviations from the experimental data can be largely traced
back to the slow convergence of the chiral expansion of the nucleon
form factors \cite{Kubis:2000zd,Bauer:2012pv}. Indeed, substituting the phenomenological
parametrization of the nucleon form factors from Ref.~\cite{Belushkin:2006qa}, our
predictions  for $F_C (q^2)$, $F_M (q^2)$ and $F_Q (q^2)$ are in an
excellent agreement with the data even at large values of $q$.
For the deuteron magnetic and quadrupole moments we obtain the values
of $\mu_M^{\rm LO} = 0.826(e/(2m))$ and $\mu_Q^{\rm LO}=0.271$ fm$^2$
in good agreement with the experimental numbers of $\mu_M=0.85741 (e/(2
m))$ and $\mu_Q=0.2859\ {\rm fm}^2$.  The observed deviations of the
order of $\sim 3.5\%$ for the magnetic moment and $\sim 5\%$ for the
quadrupole moment are consistent with
the expected size of higher-order corrections due to the two-nucleon
currents \cite{Kolling:2012cs,Piarulli:2012bn}.

It is also instructive to compare the results in the EFT with and
without explicit pions, see Fig.~\ref{fig:2}.  The pionfull approach
yields a clearly superior description of the Coulomb  and magnetic form
factors.  This is consistent with the observation that the one-pion exchange
potential plays a very important role in the $^3S_1$-$^3D_1$ channel
as witnessed e.g.~by the low-energy theorems, see
Ref.~\cite{Epelbaum:2012ua}.
Notice that the
quadrupole form factor vanishes at LO in the pionless approach since
the deuteron in this case does not have a D-state component.
Finally, we emphasize that our pionless results agree well with the
ones obtained 
within the non-relativistic
framework in Ref.~\cite{Kaplan:1998sz}.

To summarize, in the present work we calculated the electromagnetic
form factors of the deuteron at LO in an EFT using the renormalizable
approach of Ref.~\cite{Epelbaum:2012ua}.  Following
Ref.~\cite{Kaplan:1998sz}, we introduced an interpolating field for
the deuteron and calculated the form factors by applying the LSZ
reduction formalism to the three-point correlation function.
We worked out a set of integral equations (without making use of
partial wave decomposition), which are renormalizable at LO, i.e. all
ultraviolet divergences  are absorbable into a redefinition of the
parameters of the LO potential. Our parameter-free and
cutoff-independent predictions for the electromagnetic form factors of
the deuteron at LO are in very good agreement with the experimental
data.

\section*{Acknowledgments}

 MRS thanks S.~Pastore for useful discussions on electromagnetic current operators.
This work was supported in part by Georgian Shota Rustaveli National
Science Foundation (grant 11/31),
DFG (SFB/TR 16,
``Subnuclear Structure of Matter''), by the European
Community-Research Infrastructure Integrating Activity ``Study of
Strongly Interacting Matter'' (acronym HadronPhysics3,
Grant Agreement n. 283286) under the Seventh Framework Programme of EU,
ERC project 259218 NUCLEAREFT,
and the US Department of Energy under grant no. DE-SC0010300.


\end{document}